\documentstyle[aps,prb,floats,twocolumn]{revtex}
\input epsf
\begin{document}
\title{Cation- and vacancy-ordering in Li$_x$CoO$_2$}
\author{C. Wolverton and Alex Zunger}
\address{National Renewable Energy Laboratory, Golden, CO 80401\\}
\date{\today}
\maketitle
%
%
{\let\clearpage\relax
\twocolumn[%
\widetext\leftskip=0.10753\textwidth \rightskip\leftskip
\begin{abstract}
Using a combination of 
first-principles total energies, a cluster expansion
technique, and Monte Carlo simulations,
we have studied the Li/Co ordering in LiCoO$_2$ and
Li-vacancy/Co ordering in
the $\Box$CoO$_2$. 
We find: 
(i) A ground state
search of the space of substitutional
cation configurations yields
the CuPt structure as the lowest-energy 
state in the octahedral
system LiCoO$_2$ (and $\Box$CoO$_2$), 
in agreement with the experimentally observed phase.  
(ii) Finite temperature calculations predict that the solid-state
order-disorder transitions for LiCoO$_2$ and $\Box$CoO$_2$ occur 
at temperatures 
($\sim$5100 K and $\sim$4400 K, respectively) much higher than melting, 
thus making these transitions experimentally inaccessible. 
(iii) The energy of the reaction
$E_{\rm tot}(\sigma,{\rm Li}{\rm Co}{\rm O}_2) -
E_{\rm tot}(\sigma,{\Box}{\rm Co}{\rm O}_2) -
E_{\rm tot}({\rm Li, bcc})$
gives the average battery voltage $\overline{V}$
of a Li$_x$CoO$_2$/Li cell for
the cathode in the structure $\sigma$.
Searching the space of configurations $\sigma$ for large average
voltages, we find that $\sigma$=CuPt (a monolayer $\langle 111 \rangle$
superlattice) has a high voltage ($\overline{V}$=3.78 V), but
that this could be increased by cation randomization 
($\overline{V}$=3.99 V), partial disordering
($\overline{V}$=3.86 V), or by forming a 2-layer Li$_2$Co$_2$O$_4$
superlattice along $\langle 111 \rangle$
($\overline{V}$=4.90 V).

\end{abstract}
\vspace{11pt}
\pacs{PACS numbers: 61.66.Dk, 61.66.Fn, 64.70.Kb, 71.15.Ap, 81.30.Hd}

]}
\narrowtext

\section{Introduction}

%
%
\begin{figure*}[htb]
\hbox to \hsize{\centerline{
\epsfxsize=0.90\hsize\hfil\epsfbox{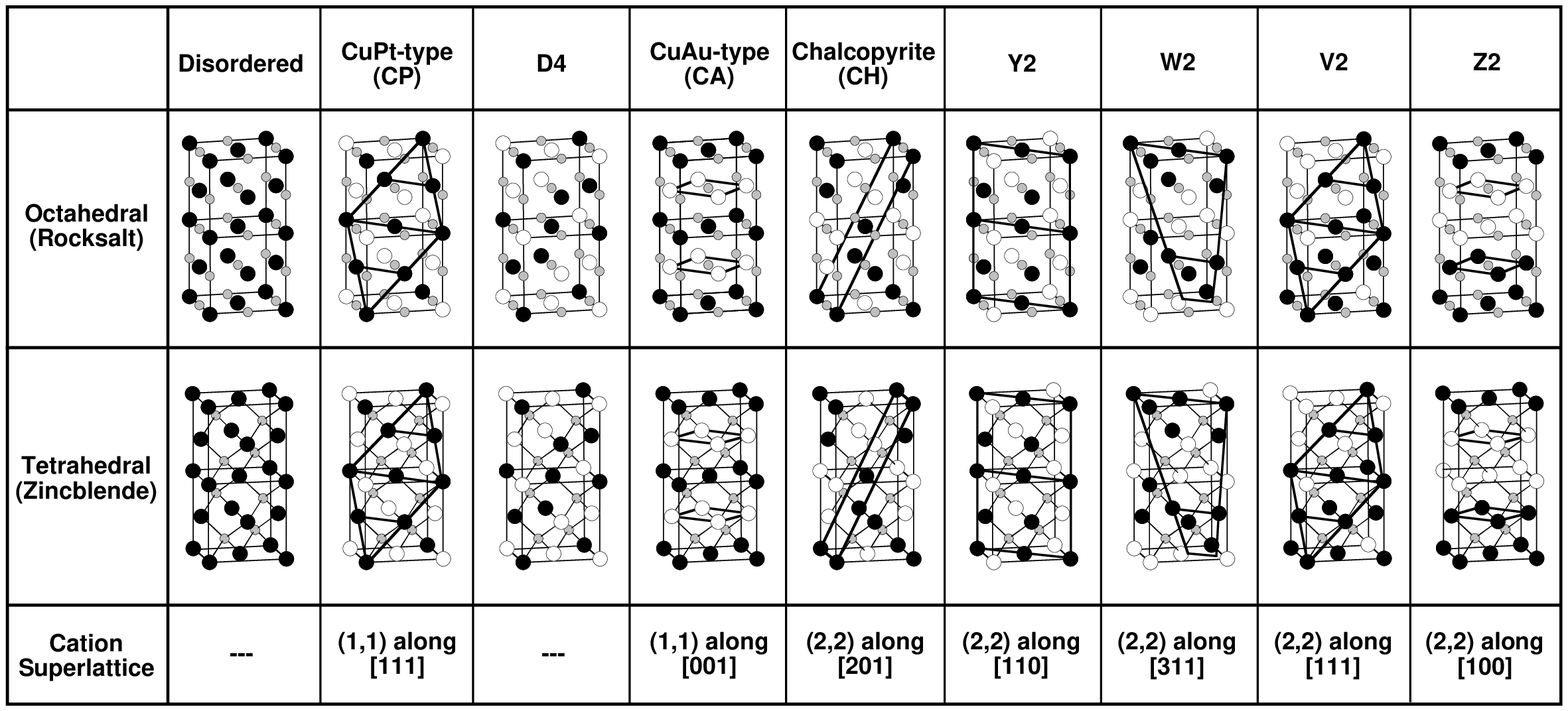}\hfil}}
\nobreak\bigskip
\caption{Cation arrangements in tetrahedral (semiconductor) and
octahedral (oxide) networks.  
The black and white atoms represent the cations, while
the grey atoms are the anions.
Shown are the names of the
cation configurations, the structure itself (bold lines indicate
superlattice planes), and the equivalent superlattice of cations.
Note:  Some of the structures are shown as only a portion of the
complete unit cell.}
\label{structures.fig}
\end{figure*}

Much like the $ABC_2$ semiconductors ($A,B$ = Al,Ga,In 
and $C$ = N,P,As, or Sb)
exhibit cation ordering in a tetrahedrally-coordinated
network \cite{Zunger.review}, 
the Li$M$O$_2$ oxides ($M$=3$d$ transition metal) \cite{Hewston,Li94}
form a similar series of
structures based on the octahedrally-coordinated network
with anions (O) on
one fcc sublattice and cations (Li and $M$) on the other
(Fig. \ref{structures.fig}).
Cation arrangements in isovalent (III-III-V) or heterovalent (I-III-VI)
semiconductor alloys have been 
observed \cite{Zunger.review} in
the disordered, CuAu-type (CA), CuPt-type (CP), and chalcopyrite (CH)
structures 
(bottom row of Fig. \ref{structures.fig}), 
while cation arrangements in the oxides have been 
observed \cite{Hewston,Li94} in 
the disordered, CP, CH, D4, and Y2 
structures (top row of Fig. \ref{structures.fig}). 
{\em Ab-initio} total energy calculations \cite{Zunger.review} have shown
that in the tetrahedrally-coordinated
III-V semiconductor alloys, the CuPt
structure is the least stable [due to the fact that it represents
a stacking along the elastically hard (111) direction], while
the chalcopyrite structure is most stable (it possesses both the
lowest electrostatic and strain energies).
Similar studies have been performed for the
octahedrally-coordinated networks of 
the spin alloy Mn$^{\uparrow}$S-Mn$^{\downarrow}$S 
and the
lead chalcogenides. \cite{Wei.oct}
In this paper,
we examine the energetics and thermodynamics of cation ordering
tendencies in the octahedral LiCoO$_2$ oxide, and  
compare to the tetrahedral semiconductor case which is well studied.
The LiCoO$_2$ compound
is used as a cathode material in rechargeable 
Li batteries. \cite{Mizushima,Julien94,Gummow92,Gummow93,Rossen93,Antaya93,Amatucci,Reimers92,Ohzuku93,Reimers93}
When Li is de-intercalated from the compound, it creates
a vacancy (denoted $\Box$) that can be positioned in different
lattice locations.
Hence, we will examine not only 
(a) the Li/Co cation ordering (different sites for Li and Co)
properties of LiCoO$_2$ ($x_{\rm Li}$=1), but also 
(b) the vacancy/Co ordering
(different sites for $\Box$ and Co) in 
$\Box$CoO$_2$ ($x_{\rm Li}$=0).
A third type of ordering in these materials, vacancy/Li 
ordering in Li$_x\Box_{1-x}$CoO$_2$ ($0 \leq x_{\rm Li} \leq 1$), 
is not treated here.

Our calculation proceeds in three steps:
(1) {\em Total energy calculations:}
We calculate the $T$=0
total energy of a set of (not necessarily stable)
ordered structures via
the full potential, all-electron
linearized augmented plane wave method (LAPW) 
\cite{singh,wei}
with all
atomic positions fully relaxed via quantum mechanical 
forces.
We then map those energies onto a
(2) {\em cluster expansion (CE)}. 
\cite{deFontaine79,Kanamori77,Sanchez84,Ducastelle91,deFontaine94,Zunger94}
This expansion is a generalized Ising-like expression
for the energy of an {\em arbitrary} substitutional
cation arrangement.  Once the coefficients of the expansion are
known, the Ising-like expression may be easily evaluated 
for any cation configuration.
Thus, one can calculate (via first-principles) the total energy
of {\em a few} cation arrangements, but then effectively search
the space of $2^N$ configurations 
(where $N$ is typically $\lesssim10^4$).
Specifically, the cluster expansion may be used to search
the entire configurational space for stable ground state
structures, where one can obtain low energy, but otherwise
unsuspected states (i.e., states which are not included in 
the set of calculated total energies).
Having obtained such a general and computationally simple
parameterization of the configuration energy, we subject it to
(3) {\em Monte Carlo simulated annealing}
\cite{Binder}
to extend first-principles
calculations (at zero temperature) to finite-temperatures,
thus obtaining order-disorder transition temperatures and
thermodynamic functions.

We find for LiCoO$_2$ the following:

(a) A ground state search of the space of substitutional
cation configurations yields
the CuPt structure as the ground state in the octahedral
LiCoO$_2$ system, in agreement with the well-established 
experimentally observed phase.  \cite{Mizushima}  
We find that this result holds even if the CuPt structure
is not included in the set of energies used to fit the
CE parameterization.
The CuPt cation structure is the least stable bulk structure in 
tetrahedral $ABC_2$.


(b) Finite temperature calculations predict that the solid-state
order-disorder transition for LiCoO$_2$ occurs at temperatures 
($\sim$5100 K) much higher than melting, thus making this transition
experimentally inaccessible.  In contrast, order-disorder transitions
in isovalent tetrahedral $ABC_2$ systems 
are $\lesssim$ 1000 K. \cite{Zunger.review}
The addition of Li vacancies lowers
this transition to $\sim$4400 K; but, this transition
temperature is still too high to be observed.  Thus, the finite
temperature calculations demonstrate that the observed
disordered (rocksalt) phase of LiCoO$_2$ is not thermodynamically
stable, but is only stabilized kinetically.

(c) The energy of intercalation reaction
$E_{\rm tot}(\sigma,{\rm Li}{\rm Co}{\rm O}_2) -
E_{\rm tot}(\sigma,{\Box}{\rm Co}{\rm O}_2) -
E_{\rm tot}({\rm Li, bcc})$
gives the average battery voltage $\overline{V}$
of a Li$_x$CoO$_2$/Li cell for
the cathode in the structure $\sigma$, \cite{Julien94}
thus providing a means for prediction of battery
intercalation voltages from first-principles energetics.
Searching the space of configurations $\sigma$ for large average
voltages, we find that $\sigma$=CuPt [a monolayer $\langle 111 \rangle$
superlattice] has a high voltage ($\overline{V}$=3.78 V), but
that this could be increased by cation randomization
($\overline{V}$=3.99 V), partial disordering
($\overline{V}$=3.86 V), or by forming a 2-layer Li$_2$Co$_2$O$_4$
superlattice along $\langle 111 \rangle$
($\overline{V}$=4.90 V).

(d) Ordered cation arrangements in LiCoO$_2$ 
are {\em stable}, similar to the heterovalent
tetrahedral I-III-VI$_2$ (e.g., CuInSe$_2$) 
systems \cite{Wei92}, but is
the opposite situation from the 
isovalent tetrahedral III-V systems such as GaInP$_2$, 
in which bulk ordered compounds are {\em unstable}.  

\begin{figure}[tb]
\hbox to \hsize{\centerline{
\epsfxsize=0.95\hsize\hfil\epsfbox{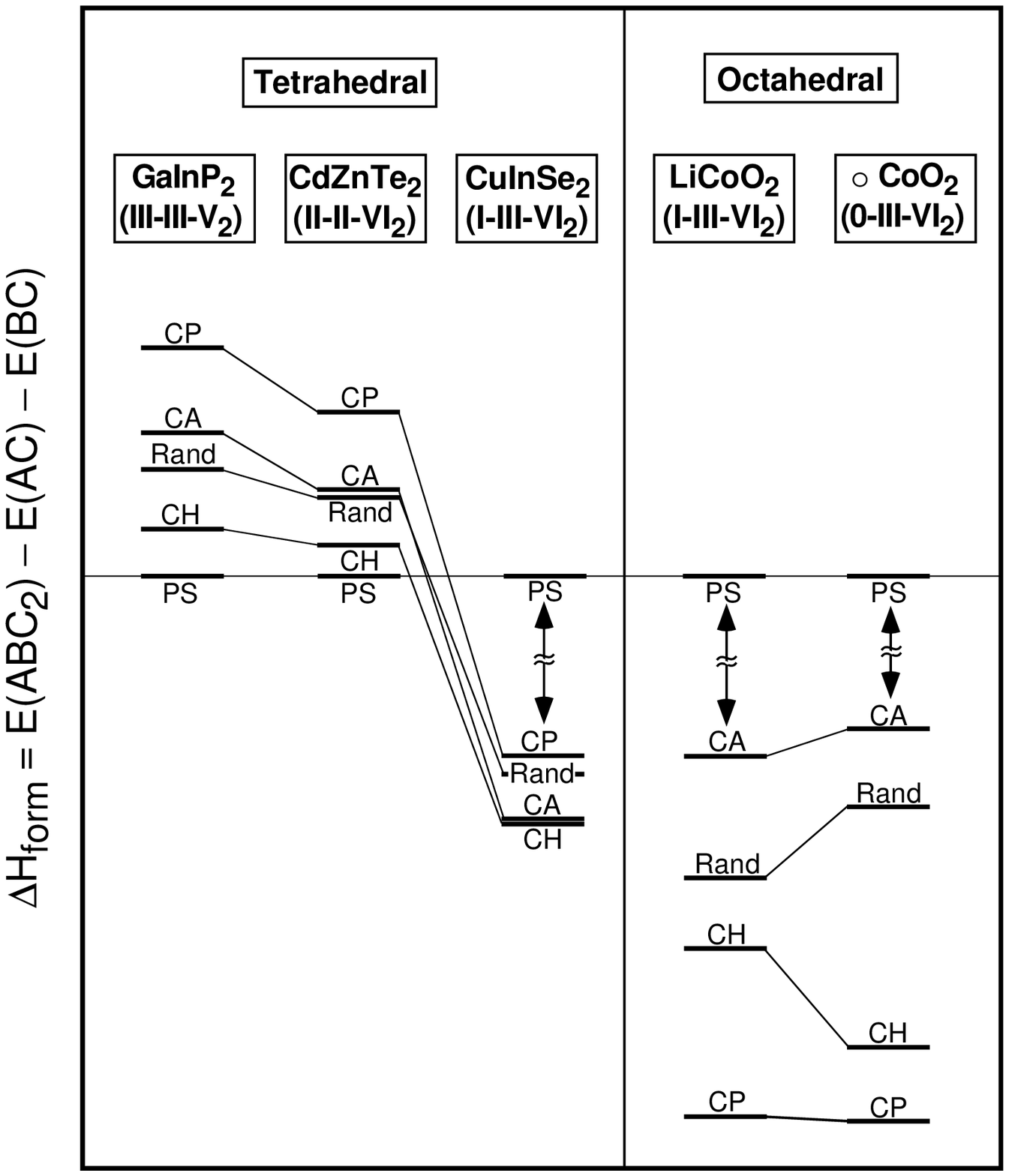}\hfil}}
\nobreak\bigskip
\caption{Formation energies of cation ordering in tetrahedral
and octahedral ABC$_2$ compounds in structures shown in 
Fig. \protect\ref{structures.fig}.  For the GaInP$_2$ and
CdZnTe$_2$ compounds, the energy scale was multiplied by
5 for visual clarity.  The GaInP$_2$, CdZnTe$_2$, and 
CuInSe$_2$ energetics were taken from Refs.
\protect\onlinecite{Silverman95,Wei90,Wei92}, respectively.
The LiCoO$_2$ and $\Box$CoO$_2$ energetics are from
the present work.  ``PS'' represents the energy of
a phase-separated mixture of AC+BC rocksalt (zincblende)
binaries in the octahedral (tetrahedral) systems.
``Rand'' is the energy of a phase in which cations
are distributed randomly (i.e., with no correlations)
on their sublattice.}
\label{tet.vs.oct}
\end{figure}

(e) The relative order of structural energies in the
octahedral LiCoO$_2$ system is quite different from the
tetrahedral cases:  $E$(CuPt) $<$ $E$(CH) $<$ $E$ (CA) in
both LiCoO$_2$ and $\Box$CoO$_2$, 
compared with $E$(CH) $<$ $E$(CA) $<$ $E$ (CuPt), universally
found in the lattice-mismatched
tetrahedral systems (Fig. \ref{tet.vs.oct}).

%
%

\section{Methods of Calculation}

%
%
\begin{table*}[htb]
\caption{Lattice averaged spin products $\overline{\Pi}_f(\sigma)$
(for a few figures $f$)
of the cation configurations $\sigma$
shown in Fig. \protect\ref{structures.fig}.}
\label{table.pis}
\begin{tabular}{ccc|cccccccccc}
&&&\multicolumn{9}{c}{$\overline{\Pi}_f(\sigma)$}\\
Interaction&Figure&$D_f$&AC+BC&CP&D4&Y2&CH&CA&W2&V2&Z2&Random\\
&$f$&&&&&&&&&&&$x$=1/2\\
\tableline
$J_0$&Empty      &1 &1&1 &1 &1     &1   &1   &1   &1   &1   &1\\
$J_1$&Point      &0 &0&0 &0 &0     &0   &0   &0   &0   &0   &0\\
$J_2$&NN Pair    &6 &1&0 &0 &0     &-1/3&-1/3&-1/6&1/2 & 1/3&0\\
$K_2$&2NN Pair   &3 &1&-1&-1&-1/3  &1/3 &1   &0   &0   & 1/3&0\\
$L_2$&3NN Pair   &12&1&0 &0 &0     &1/3 &-1/3&0   &0   &-1/3&0\\
$M_2$&4NN Pair   &6 &1&1 &1 &-1/3  &-1/3&1   &0   &0   &-1/3&0\\
$N_2$&5NN Pair   &12&1&0 &0 &0     &-1/3&-1/3&1/6 &-1/2& 1/3&0\\
$O_2$&6NN Pair   &4 &1&-1&-1&1     &-1  &1   &0   &0   &-1  &0\\
$P_2$&7NN Pair   &24&1&0 & 0&0     &1/3 &-1/3&0   &0   &-1/3&0\\
$Q_2$&8NN Pair   &3 &1&1 & 1&1     &1   &1   &-1  &-1  &   1&0\\
$R_2$&9NN Pair   &6 &1&0 & 0&0     &-1/3&-1/3&-1/6&1/2 & 1/3&0\\
$S_2$&10NN Pair  &12&1&0 & 0&0     &-1/3&-1/3&1/6 &-1/2& 1/3&0\\
$J_4$&Tetrahedron&2 &1&-1&1 &-1    &1   &1   &0   &0   & 1/3&0\\
\end{tabular}
\end{table*}

We use the cluster expansion (CE) technique, 
\cite{deFontaine79,Sanchez84,Ducastelle91,deFontaine94,Zunger94} 
which consists of an Ising-like expression
in which each cation is associated with the site of an ideal 
lattice (fcc, in this case),
and the pseudo-spin
variable $S_i$ is given the value $+$1($-1$) if an $A(B)$ atom
is assigned to site $i$.  
Within this description, the energy of {\em any} 
configuration $\sigma$ of cations 
can be written as:\cite{Sanchez84}
\begin{equation}
\label{ce}
E_{\rm CE}(\sigma) = \sum_{f} D_f J_f \overline{\Pi}_f(\sigma),
\end{equation}
where 
$f$ is a figure 
comprised of several
lattice sites (pairs, triplets, etc.), $D_f$ is the number of figures
per lattice site, $J_f$ is the Ising-like interaction for the figure
$f$, and $\overline{\Pi}_f$ is a function defined as a product over the
figure $f$ of
the variables $S_i$, averaged over all symmetry equivalent figures of
lattice sites.  
This expression incorporates the effects of
atomic relaxation (indeed, one does not require that the cations
sit precisely at the ideal lattice positions, but merely that there is a 
one-to-one correspondence between lattice sites and atomic positions).
\cite{Ceder93}
We determine $\{J_f\}$ by fitting $E_{\rm CE}(\sigma)$ of $N_{\sigma}$
structures to LDA total energies $E_{\rm LDA}(\sigma)$, given the
matrices $\{\overline{\Pi}_f(\sigma)\}$ for these structures.
Table \ref{table.pis} gives the values of the lattice-averaged
spin products $\overline{\Pi}_f(\sigma)$ 
and degeneracies $D_f$ for the
structures in Fig. \ref{structures.fig}.  
The values of $\overline{\Pi}_f(\sigma)$ sometimes
take on interesting degeneracies: For example, 
CA and CH differ only by $\overline{\Pi}_f(\sigma)$ for
figures beyond the nearest neighbor, and
the CuPt and D4 structures have \cite{Li94,Lu91} 
equivalent pair correlation functions 
for {\em all} pair separations.  
Also, all odd-body correlation functions are zero
for any structure which possesses $A\rightarrow B$
(or $\hat{S_i} \rightarrow -\hat{S_i})$ symmetry, such as all
those shown in Fig. \ref{structures.fig}
This means that CuPt and D4 possess not only equivalent
pair correlations, but three-body correlations as well.
Thus, in terms of Eq. (\ref{ce}), all terms in the expansion 
which correspond to one-, two-, and three-body figures 
do not distinguish between CuPt and D4, and thus
the first cluster correlation which can break the
degeneracy between these two structures is in the four-body
terms.
In spite of the similarities between the CuPt and D4
structures, the former is a superlattice along the [111]
direction and hence is rhombohedral, while the latter
is not a superlattice and is cubic 
(see Fig. \ref{structures.fig}).  Thus, the CuPt 
structure has one extra structural degree of freedom 
(namely a $c/a$ ratio) that the D4 structure 
does not have. \cite{note.cupt.d4}

%
%
\begin{table*}[htb]
\caption{Predicted structural information of 
observed LiCoO$_2$ phases. Where available, experimental
data and other calculated results are shown.
CuPt - ABC and CuPt - AAA
refer to CuPt configurations of cations (alternately
stacked Li/Co or $\Box$/Co close-packed layers) in
ABC... or AAA... type stackings, respectively.
For $\Box$CoO$_2$, CuPt - ABC and CuPt - AAA
are isostructural with CdCl$_2$ and CdI$_2$,
respectively.
Bulk moduli for LiCoO$_2$ (CuPt)
and $\Box$CoO$_2$ (CuPt) were calculated (present work)
to be 2.4 and 2.8 Mbar, respectively.}
\label{lattice.const}
\begin{tabular}{ccddddd}
Compound&Method&$a$ (\AA)&$c$ (\AA)&Li-O (\AA)&Co-O (\AA)
&$V_{\rm eq}$ (\AA$^3$)\\
\tableline
\\
LiCoO$_2$ (CuPt - ABC)&
Expt. \tablenotemark[1]
	    &2.82  & 14.04   & 2.07 & 1.94 & 32.23   \\
&FLAPW (present work)
	    &2.81  & 13.60   & 2.08 & 1.90 & 31.2    \\
&Pseudopot. \tablenotemark[2]
	    &2.93  & 13.2    & 2.10 & 1.96 & 32.71   \\
\\
$\Box$CoO$_2$ (CuPt - ABC)&
FLAPW (present work)      
	    &2.78  & 12.13   &      & 1.85 & 26.9    \\
&Pseudopot. \tablenotemark[2]
	    &2.88  & 12.26   &      & 1.90 & 29.36   \\
\\
LiCoO$_2$ (CuPt - AAA)&
FLAPW (present work)
	    &2.79  & 4.74    & 2.11 & 1.90 & 32.0    \\
\\
$\Box$CoO$_2$ (CuPt - AAA)&
Expt. \tablenotemark[3]
	    &2.822 &  4.29   &      & 1.91 & 29.6    \\
&FLAPW (present work)       
	    &2.80  &  4.01   &      & 1.85 & 27.1    \\
\\
LiCoO$_2$ (D4)&
Expt. \tablenotemark[4]
	    &8.002 &         & 2.06 & 1.95 & 32.0    \\
&FLAPW (present work)
	    &7.90  &         & 2.05 & 1.91 & 30.7    \\
\\
$\Box$CoO$_2$ (D4)&
FLAPW (present work)
	    &3.85  &         &      & 1.85 & 28.5    \\
\end{tabular}
\tablenotetext[1]{Ref. \onlinecite{Ohzuku93}}
\tablenotetext[2]{Ref. \onlinecite{Ceder97a}}
\tablenotetext[3]{Ref. \onlinecite{Amatucci}}
\tablenotetext[4]{Ref. \onlinecite{Antayab}}
\end{table*}

We use $N_{\sigma} = 8$ configurations 
in the fitting procedure. These are shown 
in Fig. \ref{structures.fig}.
The choice of end-point configurations requires
some discussion.  The nominal end point configurations, 
LiO and CoO in the NaCl structure, do not obey the
octet rule, as LiO has 7 valence electrons/formula unit,
while CoO has (in addition to its filled $t_{2g}$ shell)
9 valence electrons/formula.  As a result, these
nominal structures have a very high energy.  In
the 1:1 structures (LiO)$_n$(CoO)$_n$, 
an electron will move from each CoO unit to
fill the hole in the LiO unit, thus creating normal
octet bonds.  These ``charge compensated'' end-point
compounds (LiO)$^*$ and (CoO)$^*$ will have a lower
energy than the nominal LiO and CoO.  Our calculations
thus consider only charge compensated structures.
Using the procedure of Wei {\em et al.} \cite{Wei92} in treating
heterovalent alloys, the 
conventional, high-energy ``end-point''
compounds LiO+CoO are not included in the CE because 
they are not charge compensated.  
Our CE could be used to {\em predict} the energies of
(LiO)$^*$ + (CoO)$^*$, and we will see that this energy
is indeed lower than that of nominal LiO+CoO.
We only include the eight (LiO)$_n$(CoO)$_n$ compounds shown in
Fig. \ref{structures.fig} in our fit.
These cation orderings correspond to both the observed structures 
in the Li$M$O$_2$ series (CuPt, D4, Y2, and CH)
as well as other cation arrangements not observed in this series
(CuAu,W2,V2, and Z2).
For substitutional ordering problems, it is possible to 
enumerate all configurations up to a given unit cell size. \cite{Guima91}
The set of cation configurations considered here includes
all of the possible equiatomic structures with unit
cell size up to eight atoms. 

The set of twelve figures $f$ retained in the expansion
is the ``empty'' figure, the first ten neighbor pairs,
and the nearest neighbor tetrahedron.
In fitting the LDA total energies to the cluster expansion,
we include a Lagrange multiplier with the constraint that
the pair interactions should be as smooth as possible
in reciprocal space.  This technique (more fully explained
in Ref. \onlinecite{Laks92}) allows one to retain more
figures in the expansion than total energies, and also
requires the pair interactions to be as convergent as
possible in real space.
Although more sophisticated versions of the cluster expansion
approach \cite{Laks92} are available when one requires
extreme accuracy and has access to a large database of
structural energies, we use the simple real-space expansion
of Eq. (\ref{ce}) with the Lagrange multiplier. 
The use of this simple form is predicated
on the assumption that if we specialize to fixed composition
(e.g., either $x_{\rm Li}$=1 or $x_{\rm Li}$=0)
the expansion converges quickly with
a small number of terms.   
For our CE, the error in
the input energies used in the fit is a negligible amount,
$<$ 1 meV/formula unit.
To obtain some idea of
the errors involved in the CE predictions,
we have removed some structures and figures from the 
fitting process, and examined the resulting errors:
Removing the CuPt (CH) structure and the four-body tetrahedron
figure from the fit
produces an 11 (49) meV/formula unit error in the energy of CuPt (CH),
negligible changes in the other fitted energies, and
an energy of the random cation arrangement
which changes by only 1 (7) meV/formula unit.
The magnitude of these errors is quite small in terms of
the energetic scales of cation ordering 
($\sim$1000-2000 meV/formula unit) and
Li intercalation ($\sim$4000 meV/formula unit).

The expression of Eq. (\ref{ce}) can be applied to different
ordering problems, with a separate expansion constructed for
each situation.  Here, we construct three separate
cluster expansions to describe three different types of 
structural energetics:

(a) Formation enthalpies for different Li/Co arrangements $\sigma$
on the fcc lattice:
\begin{eqnarray}
\label{deltah.licoo2}
\Delta H_f&&(\sigma,\underline{\rm Li}\underline{\rm Co}{\rm O}_2) 
= 
E_{\rm tot}(\sigma,\underline{\rm Li}\underline{\rm Co}{\rm O}_2) 
- E_{\rm tot}({\rm LiO},B1) 
\nonumber \\
&&- E_{\rm tot}({\rm CoO},B1)
\end{eqnarray}
where the last two terms refer to LiO and CoO in the NaCl ($B1$)
structure with lattice constants obtained by minimizing the 
respective total energies with respect to hydrostatic deformation.
The resulting CE will reveal Li/Co ordering tendencies at 
$x_{\rm Li}$=1.

(b) Formation enthalpies for different $\Box$/Co arrangements $\sigma$
on the fcc lattice:
\begin{eqnarray}
\label{deltah.coo2}
\Delta H_f&&(\sigma,\underline{\Box}\underline{\rm Co}{\rm O}_2) 
= 
E_{\rm tot}(\sigma,\underline{\Box}\underline{\rm Co}{\rm O}_2) +
E_{\rm tot}({\rm Li,bcc}) \nonumber \\
&& - E_{\rm tot}({\rm LiO},B1) - E_{\rm tot}({\rm CoO},B1)
\end{eqnarray}
where $E_{\rm tot}({\rm Li,bcc})$ is the total energy of Li in
the bcc structure with lattice constant obtained from total
energy minimization.
The resulting CE will reveal $\Box$/Co ordering tendencies at 
$x_{\rm Li}$=0.

(c) The Li battery intercalation reaction energy
for different Li/Co (and $\Box$/Co) arrangements on the fcc lattice:
\begin{eqnarray}
\label{react}
\Delta H_{\rm react}&&(\sigma) = 
E_{\rm tot}(\sigma,\underline{\rm Li}\underline{\rm Co}{\rm O}_2) 
- E_{\rm tot}(\sigma,\underline{\Box}\underline{\rm Co}{\rm O}_2) 
\nonumber \\
&&- E_{\rm tot}({\rm Li, bcc})
\end{eqnarray}
$\Delta H_{\rm react}$ is the energy gained upon complete
de-intercalation of Li from LiCoO$_2$, relative to Li metal,
and is simply the difference between 
Eq. (\ref{deltah.licoo2}) and Eq. (\ref{deltah.coo2}).
If one assumes that the Li is removed without a change of the 
cathode structure $\sigma$ (a topotactic reaction),
\begin{equation}
{\rm Li}{\rm CoO}_2(\sigma) \rightarrow {\rm Li}_{1-x}{\rm CoO}_2(\sigma) 
+ x{\rm Li}^+ + x{\rm e}^-,
\end{equation}
then (see, e.g., Ref. \onlinecite{Julien94}) the
reaction energy $\Delta H_{\rm react}$ of Eq. (\ref{react})
is equal to the integral
of the (zero temperature and pressure) open circuit voltage $V$
of a Li$_x$CoO$_2$/Li cell between Li compositions
$x_{\rm Li}$=0 and $x_{\rm Li}$=1:
\begin{equation}
\Delta H_{\rm react}(\sigma) = -F\int_0^1 dx V(\sigma,x) = -\overline{V}(\sigma)
\end{equation}
where $F$ is the Faraday constant.
Hence, $|\Delta H_{\rm react}|$ is simply the intercalation
voltage {\em averaged} over Li composition.
We note that the intercalation voltage calculated in this manner
is a bulk, thermodynamic quantity and does not contain
contributions from the cathode surface or from kinetic
phenomena.

The total energies needed for 
Eqs. (\ref{deltah.licoo2})-(\ref{react}) have been obtained
using the first-principles 
full-potential LAPW \cite{singh} method. 
In the LAPW calculations, we used the
exchange correlation of Ceperley and Alder
as parameterized by Perdew and Zunger. \cite{ceperley}
LAPW sphere radii were chosen to be 2.0, 2.0, and 1.3 a.u. for 
Li, Co, and O, respectively. 
A well converged basis set was used, corresponding to an
energy cutoff of 25.5 Ry ($RK_{\rm max}$=6.57).
Tests were performed placing the Co 3$p$ levels in a separate
semi-core energy window as opposed to treating the Co 3$p$
as a core state; negligible
differences were found, and thus the latter was used in
all the calculations described below.
Brillouin-zone integrations are performed using the
equivalent {\bf k}-point sampling
method,
using {\bf k}-points for each structure
corresponding to the same
28 (6x6x6) special {\bf k}-points 
for the fcc structure. \cite{froyen}
All total energies are optimized
with respect to volume as well as
all cell-internal and -external coordinates.
Convergence tests of the energy differences (with respect to
basis function cutoff, {\bf k}-point sampling, and muffin-tin
radii) indicate that the total energy differences are
converged to within $\sim$0.01-0.02 eV/formula unit.

Spin polarized calculations were performed for LiCoO$_2$ 
and $\Box$CoO$_2$ in
the CuPt cation arrangement in both ferromagnetic (FM) and
anti-ferromagnetic (AFM) geometries.
For LiCoO$_2$, both the FM and AFM calculations 
converged to the non-magnetic solution ($\mu_{\rm Co}$=0).  However, for
$\Box$CoO$_2$, both FM and AFM calculations showed a 
weakly magnetic solution ($\mu_{\rm Co}\sim$0.45
for both FM and AFM)
with the total energy of the FM (AFM) state 
being 14 ($\sim$0) meV/formula unit below the non-magnetic state.
Because spin polarization only has a small effect on the energy of 
these compounds, the calculations below for the energetics of 
cation ordering in LiCoO$_2$ and $\Box$CoO$_2$ are non-magnetic (NM). 
NM, FM, and AFM (with the observed
alternating [111] layers of spins) calculations were also performed
for CoO, with the AFM solution being lowest in energy, and hence
used here.  \cite{CoO}

Having obtained the coefficients $\{J_f\}$ of the CE
of Eq. (\ref{ce}) for the three types (a)-(c) of ordering
reactions, we subjected $E_{\rm CE}(\sigma)$ to a Monte Carlo
simulated annealing method for treating the configurational
thermodynamics.\cite{Binder}  A system size of 16$^3$ = 4096
atoms (with periodic boundary conditions) was used in all 
calculations.  Monte Carlo simulations were performed
in the canonical ensemble 
with the transition temperatures
being calculated from the discontinuity in the internal
energy as a function of temperature, and the ground states
determined from the simulation at a temperature
where all configurational changes proved to be energetically
unfavorable.  
This gives:  
(i) the $T$=0 K ground state structures (from a
simulation of a finite size cell initially at high temperature,
and subsequently slowly cooled to a low temperature where all
configurational changes proved to be energetically unfavorable),
(ii) the pair correlation functions or atomic short-range order
present in the disordered alloy, and 
(iii) the order-disorder transition temperature, $T_c$.  

\section{$T=0$ Formation Energies}
\label{results.t=0}

\subsection{Energetics of Li/Co ordering in LiCoO$_2$}

%
%
\begin{table*}[htb]
\caption{FLAPW calculated formation
energies (eV/formula unit)
of various cation arrangements in LiCoO$_2$
and $\Box$CoO$_2$+Li(bcc):
$\Delta H_f(\sigma,{\rm LiCoO}_2)$, $\Delta H_f(\sigma,\Box{\rm CoO}_2)$,
(formation energies of $\sigma$)
and $\Delta H_{\rm react}(\sigma)$ (average intercalation
voltage of LiCoO$_2$ relative to Li)
are defined in 
Eqs. (\protect\ref{deltah.licoo2})-(\protect\ref{react}).
$V_{\rm eq}$ is the equilibrium volume 
(\AA$^3$/formula unit)
of LiCoO$_2$,
and $\delta V$ is the change in volume upon Li extraction
[i.e., $V_{\rm eq}$(LiCoO$_2$) - $V_{\rm eq}$($\Box$CoO$_2$)].
All energies 
of various ordered, disordered, and partially-ordered
cation arrangements in LiCoO$_2$
are from cluster expansions (CE) of FLAPW energetics, 
and are described in the text.}
\label{licoo2.ener}
\begin{tabular}{cddddd}
&LiCoO$_2$&$\Box$CoO$_2$+Li(bcc)\\
Cation Structure
& $\Delta H_f(\sigma)$ 
& $\Delta H_f(\sigma)$ 
& $\Delta H_{\rm react}$ 
& $V_{\rm eq}$ 
& $\delta V$ 
\\
\tableline
CuPt    &-3.38 & +0.40  & -3.78   
& 31.3   & 4.3   \\
D4      &-3.37 & +0.54  & -3.91   
& 30.5   & 1.9   \\
Y2      &-3.07 & +0.80  & -3.87   
& 31.4   & 1.9   \\
CH      &-2.84 & +0.64  & -3.48   
& 30.8   & 3.2   \\
W2      &-2.82 & +0.94  & -3.76   
& 30.6   & 3.1   \\
CuAu    &-2.23 & +1.65  & -3.88   
& 29.5   & 3.8   \\
V2      &-2.02 & +2.88  & -4.90   
& 31.3   & 1.0   \\
Z2      &-2.13 & +2.38  & -4.51   
& 31.5   & 2.2   \\
\tableline
Random($\eta$=0,SRO=0)           &-2.68 & +1.31  & -3.99   \\
Disordered($\eta$=0,SRO$\neq$0)  &-2.95 & +0.91  & -3.86   \\
CuPt($\eta$=0.88,SRO=0)          &-3.22 & +0.60  & -3.82   \\
D4($\eta$=0.88,SRO=0)            &-3.21 & +0.71  & -3.92   \\
\end{tabular}
\end{table*}

The formation energies [Eq. (\ref{deltah.licoo2})] 
of LiCoO$_2$ in various cation arrangements 
are given in Table \ref{licoo2.ener} and
calculated structural properties are shown in
Table \ref{lattice.const}.
We note that
the D4 structure is only slightly higher in energy than 
the CuPt structure.  This competition
is interesting because LiCoO$_2$ has been
synthesized in the D4 structure by solution growth
at low temperature. 
\cite{Li94,Gummow92,Gummow93,Rossen93,Antaya93,Gummow93b,Antaya,Garcia95}
(Although there was initially some discussion in the literature
about this low-temperature synthesized phase being CuPt
with imperfect long-range order \cite{Gummow92}, it is
now established that this phase is D4 (or ``D4-like'').
\cite{Li94,Gummow93b,Rossen93,Antaya93,Antaya})
The near degeneracy of the calculated energies of the CuPt and D4
structures is simply a consequence of their identical
pair and three body correlations $\overline{\Pi}_f(\sigma)$
noted above.
We will see that the four-body interaction $J_4$ which
distinguishes these structures is quite
small, consistent with the small energy difference between 
CuPt and D4.

\subsection{Energetics of $\Box$/Co ordering in $\Box$CoO$_2$}

The formation energies [Eq. (\ref{deltah.coo2})] 
of $\Box$CoO$_2$ in various $\Box$/Co arrangements
are also given in Table \ref{licoo2.ener}.
These configurations correspond to various arrangements of Co and $\Box$.
We note the following:

(1) The relative order of energetics is similar in $\Box$CoO$_2$ as in
LiCoO$_2$.  There is only one qualitative difference: CH drops in 
energy significantly upon extraction of Li, and is lower in energy
than the Y2 structure, whereas the reverse is true for LiCoO$_2$.

(2) The separation in energy between CuPt and D4 
increases in $\Box$CoO$_2$ compared to LiCoO$_2$, due to the 
symmetry of the phases:  
Upon extraction of Li in the rhombohedral CuPt structure,
the $c/a$ ratio decreases significantly, providing a significant
source of energy lowering for $\Box$CoO$_2$ - CuPt.
D4, on the other hand, is not a layered superlattice in any direction
and has cubic symmetry.  Hence, the cell parameters of $\Box$CoO$_2$ (D4) 
cannot distort in any preferred direction, and consequently,
$\Box$CoO$_2$ (D4) does not relax as much
as CuPt.

%
%
\begin{figure}[tb]
\hbox to \hsize{\centerline{\epsfxsize=0.85\hsize\hfil\epsfbox{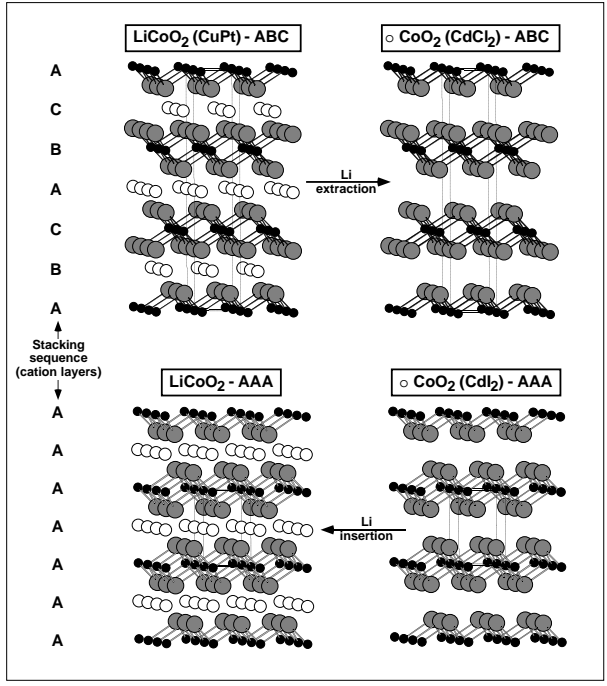}\hfil}}
\nobreak\bigskip
\caption{Different stacking arrangements of close packed cation
planes in LiCoO$_2$ and $\Box$CoO$_2$.  
The white, black and
grey atoms represent Li, Co, and O, respectively.
Clockwise from the
top left, the structures are:  
(i) The stable
LiCoO$_2$ (CuPt) phase (equivalent to the phase shown in
Fig. \protect\ref{structures.fig} but from a point of view
which emphasizes the layered nature of the compound)
with close packed cation planes 
in the ABC... fcc stacking.
(ii) The $\Box$CoO$_2$(CuPt) structure (isostructural with CdCl$_2$)
formed from extracting Li from the LiCoO$_2$ structure, 
with close packed cation planes in an ABC... stacking.  
(iii) The observed $\Box$CoO$_2$ phase ``CuPt (AAA)'' 
(isostructural with CdI$_2$)
which corresponds to an AAA... stacking of close packed cation planes.
(iv) A hypothetical LiCoO$_2$ structure ``CuPt (AAA)'', 
formed from insertion of Li into the $\Box$CoO$_2$(CdI$_2$) phase.}
\label{stacking}
\end{figure}

(3) The CuPt structure of $\Box$CoO$_2$ (isostructural 
with CdCl$_2$) has an ABC... stacking of the cation planes.
However, recent electrochemical 
measurements of Amatucci {\em et al.} \cite{Amatucci}
have succeeded in completely
de-intercalating Li from LiCoO$_2$, forming a $\Box$CoO$_2$
structure which is isostructural with CdI$_2$,
with the stacking of planes in an AAA... arrangement
(see Fig. \ref{stacking}) which we call ``CuPt (AAA)''.
These two structure are
not related to one another by substitutional degrees of freedom,
and thus are not describable by a single cluster expansion.
To examine these non-substitutional degree of freedom, we have
performed total energy calculations of $\Box$CoO$_2$
in both the CuPt and CuPt (AAA) structures (CdI$_2$).
Consistent with the observations of 
Amatucci {\em et al.} \cite{Amatucci},
we find that the $\Box$CoO$_2$ 
in the AAA stacking is lower in energy than the CuPt
structure by $\sim$0.05 eV/formula unit.

(4) We find that LiCoO$_2$ in the CuPt (AAA) structure 
(Fig. \ref{stacking})
is higher in energy
than the CuPt structure (with ABC stacking)
by $\sim$0.15 eV/formula unit,
in agreement with the fact that the
observed CuPt ground state in LiCoO$_2$ has ABC stacking.

\subsection{Effect of cation arrangement on average voltages}

Table \ref{licoo2.ener} gives the calculated reaction energies
given in Eq. (\ref{react})
for each of the cation arrangements $\sigma$ studied here.
The average voltages for all cation arrangements considered
are in the $\sim$4 V range.  In particular, the average
voltage for LiCoO$_2$ in the CuPt structure (3.78 V) is in
reasonable agreement with 
measured values (4.0-4.2 V) \cite{Mizushima,Reimers92,Ohzuku93,Amatucci} 
and pseudopotential calculations (3.75 V). \cite{Ceder97a}
Some configurations
like CH, show a marked relaxation of the $\Box$CoO$_2$ phase,
and hence show a significantly lower voltage than the
other configurations.
Thus, as also has been pointed out by previous 
authors \cite{Reimers93,Ceder97a}, we find that
first-principles calculations can provide predictions 
intercalation energies and hence, battery voltages.

An interesting aspect of the effect of cation ordering on
average voltage is that LiCoO$_2$ in the V2 structure
has a much higher average voltage than CuPt.  
This increase in voltage
is of interest because V2 is a (LiO)$_2$(CoO)$_2$ (111) 
superlattice,
whereas CuPt is a (LiO)$_1$(CoO)$_1$ (111) superlattice.
If one exchanges every other pair of cations in the CuPt layered
sequence, the V2 layering is obtained.  Thus,
V2 is just CuPt with anti-sites on two out of every four
layers.  This suggests that anti-site defects Li$_{\rm Co}$
and Co$_{\rm Li}$ in
the LiCoO$_2$ CuPt structure, while energetically very costly, 
should increase the voltage of the compound.

\section{Cluster Expansions of 
L\lowercase{i}/C\lowercase{o} and 
$\Box$/C\lowercase{o} Ordering
and $\Delta H_{\rm \lowercase{react}}$}

%
%
\begin{figure}[tb]
\hbox to \hsize{\centerline{\epsfxsize=0.95\hsize\hfil\epsfbox{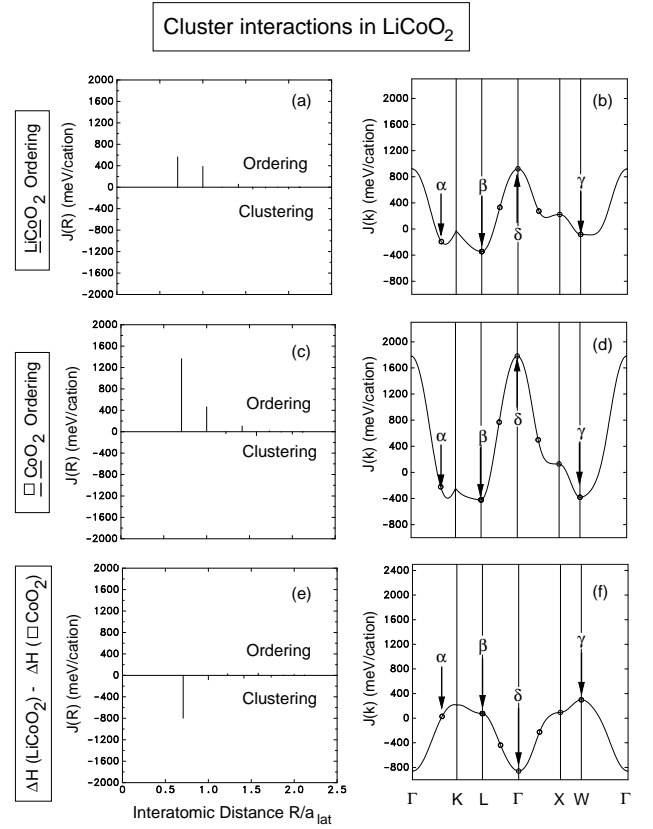}\hfil}}
\nobreak\bigskip
\caption{Pair interaction energies $J_f$ and
$J({\bf k})$ in both real (left) and reciprocal space (right).
Interactions are shown for the cluster expansions of
energies of Li/Co ordering in LiCoO$_2$ [(a) and (b)], 
energies of $\Box$/Co ordering in $\Box$CoO$_2$ [(c) and (d)], 
and average intercalation voltage in LiCoO$_2$ [(e) and (f)].
In real space, positive values of $J_f$ indicate a preferred
tendency for unlike atoms (``ordering'') and negative values 
indicate a tendency for like atoms (``clustering'').
The labels $\alpha$, $\beta$, $\gamma$, and $\delta$ indicate
the $\frac{1}{2}[110]$, $\frac{1}{2}[111]$, $\frac{1}{2}[201]$,
and $[000]$ points in reciprocal space, which correspond to
the ordering waves of the Y2, (CuPt and D4), CH, and
phase separated structures, respectively.}
\label{jofk}
\end{figure}

We can now use the set of first-principles calculated energetics 
described in Section \ref{results.t=0} to 
determine a set of interaction coefficients of the
cluster expansion [Eq. (\ref{ce})] for cation ordering.
We have constructed three cluster expansions of three different
types of ordering:
(i) Li/Co ordering in LiCoO$_2$, 
(ii) $\Box$/Co ordering in $\Box$CoO$_2$, and
(iii) cation ordering effects on the average intercalation
voltage (see below).

The pair interactions $J_f$ found from our CE fits are shown
in Fig. \ref{jofk}, both as real-space pairs $J(|{\bf R}_i-{\bf R}_j|)$
and as the lattice Fourier transform in reciprocal space $J({\bf k})$.
We see that the pair interactions in real space
are decaying with distance quite 
quickly, indicating convergence of the expansion. 
In reciprocal space, the pair interactions in Fig. \ref{jofk}
show some interesting properties:  Minima in $J({\bf k})$
indicate wavevectors where composition waves are likely to form
low energy structures.
The four body interactions $J_4$ found from our CE fits
are much smaller than
the pair interactions (e.g., $J_2$) with the ratio
$J_4/J_2$=0.004, 0.02, and 0.04 for the types of ordering
(i)-(iii) above.

For Li/Co ordering in LiCoO$_2$, 
the three {\em local} minima of $J({\bf k})$ are located at 
three wavevectors (shown by bold arrows in Fig. \ref{jofk}):  
L-point $\frac{1}{2}$(111), W-point $\frac{1}{2}$(201), and 
near the K-point $\frac{1}{2}$(110).
(Additionally, an extremely shallow minima occurs between the $\Gamma$
and $X$ points.)
These three wavevectors are the composition waves used to build
all of the structures  in the Li$M$O$_2$ series:  
CuPt and D4 [$\frac{1}{2}$(111)], CH [$\frac{1}{2}$(201)], 
and Y2 [$\frac{1}{2}$(110)].
The {\em global} minimum of $J({\bf k})$ 
occurs at the L-point [$\frac{1}{2}$(111)], 
the composition wave used to construct the CuPt and D4 structures.
Thus, we anticipate that the pair interactions $J({\bf k})$ for 
cation ordering in other Li$M$O$_2$ systems
(for other transition metals $M$) is likely to be qualitatively similar
to that of Fig. \ref{jofk} with changes in the relative minima at
these three points.
For $\Box$/Co ordering in $\Box$CoO$_2$,
the minima in $J({\bf k})$ occur at the same points as
in the case of Li/Co ordering in LiCoO$_2$, indicating 
that the relative ordering tendencies are similar in
the two systems.

For the cluster expansion of average intercalation energy, 
the minimum of $J({\bf k})$ occurs at the $\Gamma$ point,
the origin of reciprocal space
(also shown by a bold arrow in Fig. \ref{jofk}, $\delta$),
indicating that phase separation into LiO+CoO should produce
a low $\Delta H_{\rm react}$, and hence a high voltage.

Once the interactions $\{J_f\}$ are obtained, Eq. (\ref{ce}) provides
an efficient parameterization of the energy of {\em any}
configuration.
Applications of this cluster expansion
parameterization which we now discuss include a search
of configuration space (via a Monte Carlo
simulated annealing algorithm) for ground state structures,
which need not necessarily
be included in the input set, thus opening the possibility
of discovering unsuspected low energy states.
One can also perform Monte Carlo simulations at finite
temperatures to assess the thermodynamic and order-disorder
properties of the system.
Finally, one can easily calculate the energetics of disordered
and partially ordered cation arrangements in these systems.

As pointed out previously, the cluster expansions use as input
only charge compensated compounds, and therefore can be used
to predict the energies of charge compensated (LiO)$^*$ + (CoO)$^*$.
We find from our CE of LiCoO$_2$ that (LiO)$^*$ + (CoO)$^*$ is
0.79 eV/formula unit lower in energy than the nominal, 
non-charge-compensated LiO+CoO.  Similarly, the CE of
$\Box$CoO$_2$ predicts that ($\Box$O)$^*$ + (CoO)$^*$
is 0.84 eV below the non-compensated compounds.

\subsection{Ground States}

The simulated annealing algorithm finds the CuPt structure as
the low temperature state.  
In Table \ref{licoo2.ener}, 
we simply note that this structure
was the lowest in energy of the eight structures calculated by LAPW.
But, the simulated annealing prediction of the ground state
demonstrates that CuPt is also the lowest energy configuration out
of an astronomical number of possible configurations (without
symmetry, there are $\sim2^N$ possible configurations that the algorithm
could explore, where $N$=4096).  For our cluster expansion of
$\Box$CoO$_2$, the simulated annealing algorithm also
finds CuPt as the lowest energy substitutional configuration.
As we have already shown above, non-substitutional configurations
are even lower in energy for the $\Box$CoO$_2$ system (e.g., the
CdI$_2$ structure).

By combining the simulated annealing algorithm with the cluster
expansion of average voltage, one can search for the cation configuration
with {\em maximum} voltage.  This search yields a phase separated (LiO+CoO)
configuration (5.8 V).  
De-intercalating Li from this configuration would correspond
to the artificial case of: LiO $\rightarrow$ Li + $\Box$O(fcc).
It is interesting to note that other authors \cite{Ceder97a}
by very different means
have also arrived at the conclusion that this admittedly 
artificial case corresponds to the theoretical maximum voltage
in Li$M$O$_2$ compounds.

\subsection{Order-Disorder Transitions}

%
%
\begin{figure}[tb]
\hbox to \hsize{\centerline{\epsfxsize=0.90\hsize\hfil\epsfbox{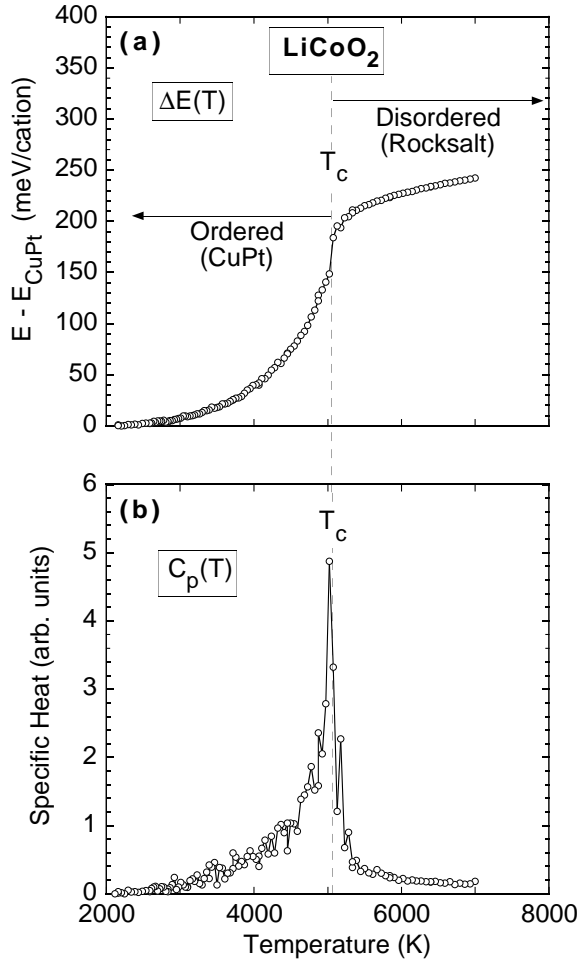}\hfil}}
\nobreak\bigskip
\caption{Monte Carlo calculated 
(a) energy relative to the $T$=0 energy of the CuPt structure and 
(b) heat capacity as functions of temperature for LiCoO$_2$.
The transition between the CuPt and disordered (rocksalt)
LiCoO$_2$ can be seen at $\sim$5100 K.}
\label{mc}
\end{figure}

For LiCoO$_2$, the order-disorder transition between
the low-temperature CuPt phase and the high-temperature
disordered phase is predicted to occur 
at $\sim$5100 K (Fig. \ref{mc}),
well above the melting point of this material.  (Note that
the calculations in this paper are all solid-state, and thus
do not consider the liquid phase.)
Antaya {\em et al.} \cite{Antaya} report a disordered rocksalt
phase of LiCoO$_2$, grown by laser ablation deposition
at 150$^{\circ}$ C, whereas growth at higher temperatures results
in either the D4 or CuPt phases.
Our calculations indicate that the observed \cite{Antaya}
disordered rocksalt phase of LiCoO$_2$
{\em is not thermodynamically stable}, but is rather
only stabilized kinetically, consistent with the fact that
the disordered phase can only be grown at low temperatures.
By performing a simulation of $\Box$/Co ordering
$\Box$CoO$_2$ at finite temperatures,
we were able to ascertain the effect of vacancies on the
order-disorder transition temperature.  Upon complete removal
of Li, the order-disorder transition of $\Box$CoO$_2$ 
drops to $\sim$4400 K, still much too high to be experimentally
accessible.  Thus, the addition of Li vacancies is not likely
to make the disordered rocksalt phase thermodynamically
accessible to experiments (although this phase is still
kinetically accessible).  

At temperatures below 2000 K, the CuPt structure is predicted to be
completely ordered.  Thus, at any growth temperatures
where thermodynamic equilibrium is achieved,
the CuPt phase should form with a long-range order (LRO)
parameter of nearly unity.
Thus, anti-site defects Li$_{\rm Co}$ or Co$_{\rm Li}$ 
are probably not formed under conditions of thermodynamic
equilibrium.
Also, since CuPt is completely ordered by 2000K, even
the D4 structure is not stabilized by thermodynamic factors
(i.e., thermal fluctuations in energy are smaller than the
CuPt-D4 energy difference for temperatures of interest). 
However, the D4 structure has been observed in low-temperature
solution grown and laser ablation-grown samples, which are
probably not equilibrium phases.

\subsection{Properties of Disordered and Partially Ordered
Cation Arrangements}

Using the CE, we can 
compute the energetics of {\em any} cation
arrangement such as random alloys
or any disordered (short-range or long-range ordered) phases.  
These are
examples of phases which are not directly accessible to 
first-principles calculations, but may be accessed via the
cluster expansion.  
We show the cluster expansion energetics of several
such phases in Table \ref{licoo2.ener}:  

{\em Random alloy}:  The perfectly random alloy
is a phase in which Li and Co atoms (or
$\Box$ and Co for $\Box$CoO$_2$) are distributed
on their cation fcc sublattice with no atom-atom
correlation between cation sites.  This corresponds to the
enthalpy as $T\rightarrow\infty$.
The energy of this
random phase is easily computed from the cluster
expansion of Eq. (\ref{ce}),
since the absence of atomic correlations leads to the
simple values $\overline{\Pi}_f = 0$, and thus the
energy of the random alloy is given by
$E_{\rm CE}({\rm Random})=J_0$.  The energies of 
random cation arrangements in LiCoO$_2$ and $\Box$CoO$_2$
are shown in Table \ref{licoo2.ener}.
The ordering energy of an ordered compound $\sigma$ is
the energy required to construct $\sigma$ from the random
cation arrangement:  
$\delta E_{\rm ord}(\sigma) = E(\sigma)-E({\rm Random})$.
From Table \ref{licoo2.ener}, we can see that for both
LiCoO$_2$ and $\Box$CoO$_2$, all ordered
cation configurations considered have $\delta E_{\rm ord}<0$,
except for CA, V2, and Z2.

{\em Partial Short-Range Order}:  
Because Antaya {\em et al.} \cite{Antaya}
report the existence of some degree of CuPt-type ordering
in their disordered phase, we have also computed 
(Table \ref{licoo2.ener}) the energetics
of a disordered rocksalt phase with some degree of
atomic short-range order (SRO).  
SRO is a finite-temperature effect, and is characterized
in real space by the 
pair correlation functions $\overline{\Pi}_{0,n} \neq 0$
for the $n$th atomic shell.
Thus, the SRO parameters, $\overline{\Pi}_{0,n}$
measure the extent to which spatial {\em correlations} 
exist in disordered alloys.
The SRO parameters used to compute the energetics of
for the first ten neighbor shells
were obtained from a Monte Carlo simulation
of the LiCoO$_2$ disordered alloy just above the order-disorder transition
(in parentheses are the values for fully
ordered CuPt or D4):  $-0.06(0.0)$, $-0.27(-1.0)$,
$+0.03(0.0)$, $+0.12(+1.0)$, $+0.02(0.0)$, $-$0.07(-1.0),
$-$0.02(0.0), $+$0.10(1.0), $-$0.01(0.0), 
and $-0.01(0.0)$.
Note that the energetic effect of SRO is to significantly
lower the energy of the random phase in both LiCoO$_2$ and
$\Box$CoO$_2$ by 0.27 and 0.40 eV/formula unit, respectively.

{\em Partial Long-Range Order}:  There have also been reports 
of long-range ordered LiCoO$_2$ (either CuPt or D4) with
small quantities of Li on the Co sites, or vice versa.  This
amounts to a CuPt or D4 phase with partial long-range order (LRO).
If the LRO parameter $\eta$=1, then all Li and Co
atoms reside completely on their own sublattice and LRO
is perfect.  However, for states of partial LRO,
$\eta<1$, and there is an amount $\frac{1-\eta}{2}$ 
of intermixing between sublattices.  For simplicity,
we assume that there are no short-range
correlations between the intermixed atoms.  In
Table \ref{licoo2.ener}, we show the energetics of
CuPt and D4 structures with LRO parameter $\eta$=0.88,
corresponding to 6\% of Li on the Co sites, and vice versa.
The LiCoO$_2$ energies of CuPt and D4
are both raised by 0.16 eV/formula unit relative to the
$\eta$=1 fully ordered phases, while the corresponding
increases for $\Box$CoO$_2$ is 0.20 and 0.17 eV/formula unit.

The cluster expansion of voltage can also be used to predict
the average voltages of configurations not directly accessible
to first-principles calculations (Table \ref{licoo2.ener}).
In particular, we see that the random alloy (3.99 V) is predicted
to have a higher average voltage than the ordered CuPt phase (3.78 V).
Since this phase has been produced by laser ablation, \cite{Antaya}
it would be interesting to measure its electrochemical properties, 
in order to compare with our predictions.
The increase in voltage due to disorder is significantly reduced
when one considers the disordered phase with SRO described 
above (3.86 V).
Thus, it is possible that experimentally,
the voltage of the disordered LiCoO$_2$
relative to CuPt could be used to indirectly determine the
amount of short-range order in the sample.
Also in Table \ref{licoo2.ener} are the voltages of 
partially long-range ordered CuPt and D4 phases 
with 6\% of the Co atoms on the Li sites ($\eta$=0.88).
Even this small amount of anti-site defects increases the voltages
of CuPt and D4 by 0.05 and 0.01 V, respectively.
Note that for either LRO or SRO, the qualitative
effect of disordering is the same:  Disorder
raises the energy of $\Box$CoO$_2$ more than LiCoO$_2$,
and thus raises the average voltage.

\section{Energetics of octahedral {\em vs.} tetrahedral $ABC_2$ networks}

We now compare our results for cation ordering
in the octahedral LiCoO$_2$ and $\Box$CoO$_2$ systems with the
well-studied cases of cation ordering in isovalent and
heterovalent tetrahedral $ABC_2$ systems.
In the (heterovalent)
octahedral Li$M$O$_2$ systems, ordered cation arrangements 
are {\em stable} with respect to LiO+$M$O. 
This ordered compound stability is qualitatively similar to
heterovalent tetrahedral systems, such as CuInSe$_2$,
where cation ordered phases are stable relative
to decomposition into CuSe+InSe zincblende
binaries.  Just as in the LiCoO$_2$ case, the
binaries correspond to compounds (rocksalt
LiO and CoO and zincblende CuSe and InSe) 
which do not satisfy the octet rule,
and hence are relatively high energy configurations.
The high energy of the constituents ``exposes''
the ordered compounds (which do satisfy
the octet rule) as stable in these I-III-VI$_2$ systems.
The stability of ordered compounds in octahedral Li$M$O$_2$ systems
is, however, in contrast to the isovalent tetrahedral
$A^{III}B^{III}C^{V}_2$ systems, in which bulk ordered compounds are
unstable with respect to AC+BC.  
In these III-III-V$_2$ systems, the phase-separated
state is the low energy ground state, and ordered compounds which
have been observed have been shown \cite{Zunger.review} 
to be a result of a combination of epitaxial strain and 
surface-reconstruction-induced ordering.

The relative order of energies of ordered compounds in the
octahedral LiCoO$_2$ and $\Box$CoO$_2$ systems 
is also quite different from the
(isovalent or heterovalent)
tetrahedral cases:  $E$(CuPt) $<$ $E$(CH) $<$ $E$ (CA) in
both LiCoO$_2$ and $\Box$CoO$_2$,
compared with $E$(CH) $<$ $E$(CA) $<$ $E$ (CuPt), universally
found in the lattice-mismatched
tetrahedral systems.
Also, in the octahedral systems,
the ordering energy $\delta E_{\rm ord}<0$ for both CuPt and CH,
while $\delta E_{\rm ord}<0$ for CH in the tetrahedral systems.

%
%
\begin{table*}[htb]
\caption{Calculated bonds lengths in various cation arrangements of
LiCoO$_2$.}
\label{bonds}
\begin{tabular}{cdd}
Cation Structure&Li-O (\AA)&Co-O (\AA)\\
\tableline
CuPt    &2.08      &1.90       \\
D4      &2.05      &1.91       \\
Y2      &1.93,2.08,2.28&1.86,1.92,1.94 \\
CH      &1.94,2.24 &1.89,1.94  \\
W2      &1.93,1.97,2.09,2.12 &1.86,1.90,1.93,1.98  \\
CuAu    &1.87,1.99 &1.87,1.99  \\
V2      &1.89,2.21 &1.84,2.04  \\
Z2      &1.99,2.04,2.35&1.76,1.96,1.99  \\
\end{tabular}
\end{table*}

The CuPt structure is preferred in octahedral networks due
to strain energy arguments:  
For an octahedral $ABC_2$ system which has distinct equilibrium
$A-C$ and $B-C$ bond lengths, the CuPt structure has the
property that the cell-internal distortion of
the anions (C) accommodates
{\em any} equilibrium $A-C$ and $B-C$ bond lengths, 
and maintains all $A-C$ bond 
lengths equal to one another (and similarly for $B-C$).  
The D4 structure also possesses this optimal structural relaxation.
This optimal bond-length accommodation is interesting in light
of the fact that the cation CuPt structure in {\em tetrahedral}
systems, when relaxed, possesses two equilibrium $A-C$ bonds (and 
similarly for $B-C$) as opposed to the single $A-C$ bond in the
octahedral case.  The distinction between the two $A-C$ bonds
in {\em tetrahedral} CuPt is due to the fact that some of the $A-C$
bonds in this structure are along the direction of cell-internal
distortion and other $A-C$ bonds are perpendicular to this direction.
Thus, when $C$ atoms are relaxed, these two types of $A-C$ bonds
adopt different bond lengths.  However, in the {\em octahedral}
CuPt structure, none of the $A-C$ bonds are either along or perpendicular
to the distortion direction of the anions, but rather all $A-C$ bonds
are at equivalent angles to this direction.  Thus, when the anions
relax, all $A-C$ bonds are distorted by equal amounts.
The calculated equilibrium bond lengths in all cation 
arrangements for LiCoO$_2$ are given
in Table \ref{bonds}.
One can see from this table that CuPt and D4 are the only
structures for which there is only one type of symmetry-inequivalent
Li-O and Co-O bond. 
Structures with Li-O and Co-O bonds
equal to one another (e.g., CH, CA) are energetically unfavorable.
In tetrahedral
systems, the configuration which possess the optimum structural
geometry, analogous to CuPt in octahedral coordination,
is the CH structure, which is the lowest energy ordered
compound in size mismatched semiconductor alloys.
This strain energy argument can also explain the relative
stability of CuPt, CH, and CA in octahedral {\em vs.} tetrahedral
systems:  Using a simple valence force field which includes
only energetic effects due to strain, one obtains the correct
order of these three structures for both octahedral and
tetrahedral systems as compared 
with LAPW. \cite{Bellaiche,Silverman95,Wei90}
One should note, however, that
in the Li$M$O$_2$ series, there are systems other than
$M$=Co which possess ground states other than CuPt, e.g.,
the CH and Y2 structures.  Thus, clearly strain-only arguments
do not explain the totality of ordering tendencies in these
compounds, as other effects must dominate in some systems.

Another distinction between the ordering tendencies of the
octahedral LiCoO$_2$ system with those of the tetrahedral
systems is in the energy scale.  In Fig. \ref{tet.vs.oct},
the energy scale of the tetrahedral systems is multiplied
by a factor of 5, and is still smaller than the octahedral
energy scale.  The difference between the energy of the
highest and lowest ordered compounds in the
isovalent tetrahedral III-III-V$_2$ systems is 
$\delta E$(CuPt-CH)$\sim$0.1 eV/formula unit, 
in the
heterovalent tetrahedral CuInSe$_2$ system it is 
$\delta E$(V2-CH)$\sim$0.7 eV/formula unit, 
whereas
this difference in the octahedral systems is 
$\delta E$(V2-CuPt)$\sim$1.4 eV/formula unit 
in the LiCoO$_2$ system and 
$\delta E$(V2-CuPt)$\sim$2.4 eV/formula unit
in $\Box$CoO$_2$.
Thus the energetic effect of cation ordering is much more
dramatic in the octahedrally coordinated networks.

\section{Summary}

Using a combination of first-principles total energy calculations,
a cluster expansion approach, and Monte Carlo simulated annealing,
we have studied the cation ordering in 
LiCoO$_2$ and $\Box$CoO$_2$, and 
compared the ordering in these heterovalent 
octahedrally coordinated systems
with previous studies of ordering in both isovalent and
heterovalent tetrahedral $ABC_2$ systems.
We find many significant differences between ordering in octahedral
and tetrahedral systems.  
In the heterovalent 
octahedral systems, ordered compounds have negative
formation energies, and are hence {\em stable}.  This
is qualitatively similar to the heterovalent tetrahedral case, 
but distinct from the isovalent 
tetrahedral semiconductors, where
ordered cation arrangements are unstable.
Also, the relative order of ordered compound energies is different 
in the octahedral systems studied here, relative to either
isovalent or heterovalent tetrahedral systems.
In particular, for both the LiCoO$_2$ and $\Box$CoO$_2$ systems, 
a simulated annealing
ground state search of the entire cation configuration space
yields the CuPt cation arrangement as the lowest energy ground state, 
whereas this structure is the {\em highest} energy configuration
in tetrahedral III-III-V$_2$ systems.
The scale of ordering energetics is dramatically different in
the LiCoO$_2$ and $\Box$CoO$_2$
octahedral systems ($\sim$1.5-2.5 eV), compared with that of
either heterovalent (e.g., $\sim$0.7 eV in CuInSe$_2$) or
isovalent tetrahedral semiconductors ($\sim$0.1 eV).  This
difference in energy scales is also reflected in the different
temperature scales of order-disorder problems in the two
types of systems:  While typical order-disorder temperatures
in isovalent or heterovalent 
tetrahedral systems are $\lesssim$1000 K, we find transition
temperatures of $\sim$5100 K and $\sim$4400 K for LiCoO$_2$
and $\Box$CoO$_2$, respectively.

Because LiCoO$_2$ is in a class of materials being studied for
use in rechargeable Li batteries, 
we have also examined the effects of 
cation ordering on Li intercalation energies 
and average voltages in Li$_x$CoO$_2$/Li cells.
Searching the space of configurations $\sigma$ for large average
voltages, we find that $\sigma$=CuPt [a monolayer $\langle 111 \rangle$
superlattice] has a high voltage ($\overline{V}$=3.78 V), but
that this could be increased by cation randomization
($\overline{V}$=3.99 V), partial disordering
($\overline{V}$=3.86 V), or by forming a 2-layer Li$_2$Co$_2$O$_4$
superlattice along $\langle 111 \rangle$
($\overline{V}$=4.90 V).

\begin{center}
{\bf Acknowledgements}
\end{center}

The authors gratefully acknowledge many helpful discussions
with Su-Huai Wei about the LAPW method and ordering in
semiconductor systems, and with Melissa Fu, David Ginley, 
Jeanne McGraw, Phil Parilla, John Perkins, and Doug Trickett
about the experimental aspects of this problem.
Work at NREL was supported by the Office of Energy Research
(OER) [Division of Materials Science of the Office of Basic Energy
Sciences (BES)], U. S.  Department of Energy, under contract No.
DE-AC36-83CH10093.


\begin{references}
\bibitem{CoO}		It is well known that CoO is a Mott insulator, 
			and the local density approximation
			incorrectly predicts that the compound 
			is a metal.
			However, in considering cation ordering, 
			LiO+CoO is simply a constant reference point 
			for these
			calculations, and so errors in the LDA CoO energy 
			will shift all formation energies of LiCoO$_2$ 
			and $\Box$CoO$_2$
			by a constant amount, and hence
			will not affect any of
			our statements about 
			ordering in LiCoO$_2$ or $\Box$CoO$_2$.
\end{references}
\end{document}